\documentclass[aps,prb,reprint,superscriptaddress,amsmath,amssymb,preprintnumbers,floatfix,frontmatterverbose,nofootinbib,nobibnotes,showkeys]{revtex4-2}
\usepackage{amsfonts,graphicx,xcolor}
\definecolor{myred}{rgb}{0.843137,0.0980392,0.12549}
\definecolor{myblue}{rgb}{0.121569,0.294118,0.627451}
\definecolor{mygreen}{rgb}{0.0,0.65,0.0}
\definecolor{mygray}{rgb}{0.45,0.45,0.45}
\definecolor{mylightblue}{rgb}{0.187,0.335,0.678}
\definecolor{myorange}{rgb}{0.952,0.423,0.093}
\usepackage[pdfstartview=FitH,colorlinks=true,linkcolor=blue,citecolor=red,urlcolor=mygreen,filecolor=mygray,anchorcolor=myorange,pdfborder={0 0 0}]{hyperref} 
\usepackage{braket}
\newcommand{\canc}[1]{{}}
\newcommand{\msc}[1]{\normalfont\textsc{#1}}
\def\re{\operatorname{Re}}
\def\im{\operatorname{Im}}
\def\Gm{\tilde{\Gamma}}
\def\Mm{\tilde{M}}
\def\Km{\tilde{K}}
\newcommand{\md}[1]{\tilde{d}_{#1}}
\def\EF{E_{\mathrm{F}}}
\def\dEFo{{\Delta}E_{\mathrm{F}}}
\newcommand{\dEF}[1]{{\Delta}E_{\mathrm{F}}^{\theta_{#1}}}
\def\vFo{\mathrm{v}_{\mathrm{F}}}
\newcommand{\vF}[1]{\mathrm{v}_{\mathrm{F}}^{\theta_{#1}}}
\def\epsM{\epsilon_{\mathrm{M}}}
\def\Eloss{E_{\mathrm{LOSS}}}
\def\cinecaurl{http://www.cineca.it/}
\def\cinecainfn{http://www.hpc.cineca.it/news/framework-collaboration-agreement-signed-between-cineca-and-infn}
\def\infnprog{INF16\_npqcd}
\begin{document}
\def\infnlnfcs{{INFN}, Gruppo collegato di Cosenza, Via P. Bucci, Cubo 31C, I-87036 Rende (CS), Italy}
\def\fisunical{{Dip. di Fisica, Universit\`{a} della Calabria}, Via P. Bucci, Cubo 30C, I-87036 Rende (CS), Italy}
\title{Plasmons in twisted bilayer graphene across dispersive and flat
bands}
\author{Antonio Palamara}
\affiliation{\fisunical}
\affiliation{\infnlnfcs}
\author{Michele Pisarra}
\affiliation{\fisunical}
\affiliation{\infnlnfcs}
\author{Antonello Sindona}
\affiliation{\fisunical}
\affiliation{\infnlnfcs}
\begin{abstract}
The dynamical dielectric response of twisted bilayer graphene is explored in large-angle dispersive-band and small-angle quasi-flat-band regimes using time-dependent density-functional theory within the random-phase approximation.
At the reference bilayer-graphene interlayer distance, weak coupling in the largest-angle structures preserves Dirac dispersions and the intrinsic $\pi$ plasmon.
Electron doping activates a two-dimensional Dirac plasmon with energies obeying approximate geometric twist-angle scaling, while acoustic-like branches remain embedded in the single-particle continuum.
The prohibitively large first-magic-angle supercell is represented by a tractable cell with its interlayer separation reduced to the angle-dependent magic distance, where four quasi-flat bands emerge around the Fermi level.
Their partial occupation produces a dispersive low-energy plasmon-like excitation without a clear dielectric zero at resonance. 
A distinct interband plasmon is instead identified, supported by transitions involving the quasi-flat manifold and neighboring high-density-of-states regions.
Band-energy rescaling places its characteristic energy in the mid-infrared range of interband collective excitations measured near the magic angle.
\end{abstract}
\maketitle
\section{Introduction\label{intro}}
Reduced dimensionality, stacking, and proximity effects make two-dimensional~(2D) materials a versatile platform for tailoring electronic states~\cite{Ren_2026}. 
Within this broad class, moir\'{e} superlattices formed by rotationally misaligned van der Waals layers occupy a distinctive position because their emergent length scale provides geometric control over bandwidths and interaction scales~\cite{Hennighausen_2021}.

A paradigmatic realization is twisted bilayer graphene~(TBG), in which a small relative rotation $\theta$ of the layers generates a moir\'{e} modulation that hybridizes their electronic states and yields a four-band low-energy manifold around the Fermi level~\cite{PhysRevB.81.245412, PhysRevB.86.125413, PhysRevB.86.155449, PhysRevLett.99.256802, PhysRevB.82.121407,doi:10.1073/pnas.1108174108}. 
At the first magic angle, $\theta\simeq1.1^\circ$, these bands become exceptionally narrow, strongly reducing the kinetic-energy scale relative to the Coulomb interaction~\cite{doi:10.1073/pnas.1108174108, PhysRevB.99.195419}. 
The resulting flat-band regime hosts correlated insulating and superconducting states, orbital magnetism, and topologically nontrivial phases~\cite{Xie2019, Kerelsky2019, Jiang2019, Choi2019, Cao2018, Cao2018a, doi:10.1126/science.aav1910, Lu2019, doi:10.1126/science.aaw3780}.

Among these phenomena, the microscopic origin of superconductivity in magic-angle TBG~(MATBG) remains unsettled.
Experiments separating superconductivity from correlated insulating order, together with studies in which environmental Coulomb screening is tuned, demonstrate a nontrivial interplay between superconductivity and electronic interactions and place strong constraints on viable pairing scenarios~\cite{Stepanov2020, Saito2020, Xiaoxue2021, Gao2026}.
Against this background, the long-standing possibility of electronically mediated superconductivity arising from dynamical Coulomb screening~\cite{PhysRevB.29.6132} has motivated theoretical studies of plasmon-mediated and mixed plasmon-phonon pairing mechanisms in TBG~\cite{PhysRevResearch.2.022040, PhysRevB.103.235401, PhysRevB.109.045404}.
 
The dynamical dielectric response is therefore relevant both to these pairing scenarios and to charge dynamics in TBG more broadly, since it determines the frequency- and momentum-dependent screened Coulomb interaction and carries signatures of collective charge excitations.
Theoretical studies have predicted weakly dispersive interband plasmons, flat-band collective modes near the magic angle, and intrinsic acoustic plasmon branches whose dispersions and damping channels differ markedly from those of monolayer and Bernal-stacked bilayer graphene~\cite{Stauber2016, PhysRevB.103.115431, PhysRevB.102.125403, PhysRevB.110.045431, doi:10.1073/pnas.1909069116}.
Near-field infrared measurements have further revealed a propagating plasmon mode at charge neutrality attributed to interband transitions between moir\'{e} minibands~\cite{Hesp2021}.
These results emphasize the importance of determining how the plasmon spectrum evolves as the underlying bands change from dispersive to quasi-flat.

Existing descriptions of these modes have relied predominantly on continuum or tight-binding approaches~\cite{Stauber2016, PhysRevB.103.115431, PhysRevB.102.125403, PhysRevB.110.045431, PhysRevB.106.155402}. 
Moving beyond such modeling to a fully atomistic first-principles response calculation entails a steep increase in computational cost. 
The number of atoms in a commensurate moir\'{e} cell grows rapidly with decreasing twist angle, reaching values of order $10^4$ near the first magic angle.
For cells of this size, obtaining the underlying Kohn-Sham~(KS) electronic structure with conventional plane-wave~(PW) density-functional theory~(DFT) is already extremely demanding, even before the dynamical response is evaluated~\cite{PhysRevB.99.195419, Palamara2025}.
Recent implementations based on optimized localized basis sets have expanded the supercell sizes accessible to DFT for structural and electronic studies~\cite{8gvw-4rgk,2m63-b51d}, whereas corresponding first-principles calculations of the momentum-dependent dielectric response for comparably large supercells remain scarce. 

The present study addresses this gap by characterizing the dielectric response and electron energy-loss spectra of selected commensurate TBG supercells using time-dependent DFT~(TDDFT) within the random-phase approximation~(RPA).
The commensurate construction retains the moir\'{e} periodicity explicitly under periodic boundary conditions, with the KS states represented in a PW basis and ground-state exchange-correlation effects described using the Perdew-Burke-Ernzerhof~(PBE) functional~\cite{PhysRevLett.77.3865}.

Calculations are performed for two commensurate configurations at the reference bilayer-graphene interlayer separation, here termed \textit{reference-distance configurations}, and for a third, smaller-angle configuration at a reduced interlayer separation chosen to access the flat-band regime.
This reduced separation is determined using the magic-distance construction established in previous work~\cite{Palamara2025}, thereby avoiding direct treatment of the prohibitively large $\theta\simeq1.1^\circ$ supercell. 
For twist angles below $6^\circ$, this construction identifies an angle-dependent reduced interlayer separation that produces a quasi-flat low-energy manifold at fixed twist angle.
The resulting constrained structure is termed the \textit{magic-distance configuration} and represents a flat-band analogue of MATBG rather than a genuine magic-angle structure at the reference interlayer separation.
Comparison of this configuration with the two reference-distance configurations allows the dispersive- and flat-band regimes to be contrasted within a common atomistic framework, with the resulting spectral differences reflecting the combined effects of twist angle and interlayer coupling.
For the reference-distance configurations, the response is surveyed from the terahertz to the ultraviolet range, whereas the magic-distance analysis focuses on the low-energy features associated with the quasi-flat manifold.

The computational implementation and the PW formulation of the dielectric response are detailed in Appendices~\ref{sec2b} and~\ref{sec3}. 
Correlation-driven symmetry breaking, self-consistent filling-dependent band reconstruction, and exchange-correlation kernels beyond RPA are outside the scope of the present analysis, as their consistent treatment within the present PW-supercell framework remains computationally impractical for the supercell sizes considered, including that of the magic-distance configuration.    
The central quantity used throughout the following analysis is the electron energy-loss function,
\begin{equation}
\Eloss(\mathbf{q},\omega)=-\im\left[\epsM(\mathbf{q},\omega)^{-1}\right],
\label{eqloss}
\end{equation}
where $\epsM$ denotes the macroscopic dielectric function evaluated within TDDFT-RPA. 
$\Eloss$ describes the dissipative response to a longitudinal perturbation that transfers in-plane momentum $\mathbf{q}$ and energy $\omega$ to the system. Within the first Born approximation, it provides the material-response factor entering the inelastic electron-scattering cross section, with additional prefactors determined by the scattering geometry and kinematics.  

Although collective modes generally produce maxima in $\Eloss$, loss peaks can also arise from single-particle electron-hole excitations and therefore do not, by themselves, establish plasmonic character.
For loss features visible in the macroscopic response, plasmonic character is therefore assessed from the dielectric function at the corresponding loss resonance. 
In the weak-damping limit, a longitudinal plasmon is associated with a zero crossing of $\re[\epsM(\mathbf{q},\omega)]$ and a sufficiently small $\im[\epsM(\mathbf{q},\omega)]$, although damping can shift the loss maximum away from the crossing~\cite{Andersen2013,Ramakrishna2021}.
A near-zero minimum of the real part accompanied by a small imaginary part is regarded here as suggestive of plasmon-like behavior, but does not by itself establish a collective mode.
If the real part neither crosses nor approaches zero near the loss resonance, the macroscopic response provides no clear dielectric evidence for a well-defined plasmon.
These signatures guide the interpretation of the reference-distance and magic-distance loss spectra.

\section{Commensurate TBG structures\label{sec2a}}
For two unstrained graphene layers, periodic TBG lattices are obtained only at the discrete \textit{commensurate} twist angles $0<\theta(m,r)<\pi/3$ satisfying 
\begin{equation}
\cos\theta(m,r)=\frac{3m^2+3mr+r^2/2}{3m^2+3mr+r^2},
\label{eq:qmr}
\end{equation} 
where $m$ and $r$ are coprime positive integers~\cite{PhysRevB.98.085435}. 
Here, the analysis is restricted to the $r=1$ family and $\theta(m,1)$ is denoted $\theta_m$.
The values $m=1$, $m=2$, and $m=7$ yield the twist angles $\theta_1=21.7868^\circ$, $\theta_2=13.1736^\circ$, and $\theta_7=4.4085^\circ$, with moir\'{e} unit cells containing $N_m=4(3m^2+3m+1)$ carbon atoms, namely $28$, $76$, and $676$ atoms, respectively. 

The $\theta_1$ and $\theta_2$ structures have the two smallest nontrivial moir\'{e} unit cells within this commensurate family and have been extensively adopted as large-angle atomistic benchmarks~\cite{Song2019, Gao2024}.
Related large-angle rotational domains have also been observed and characterized in three-dimensional nanoporous graphene~\cite{PhysRevB.111.045432}. 
All three structures are treated at the reference graphene lattice constant $a=2.46$~{\AA}.
For the $\theta_1$ and $\theta_2$ configurations, the interlayer separation is fixed at the reference value $d_0=3.349$~{\AA} of few-layer graphene and graphite. 
The $\theta_7$ lattice provides a useful compromise between a substantially reduced twist angle and the computational tractability of PW response calculations.
To access the flat-band regime, this structure is instead treated at the magic distance $\md{7}=2.587$~{\AA}, at which the four bands adjacent to the pristine Fermi level become quasi-flat~\cite{Palamara2025}. 

The three configurations analyzed are therefore $(\theta_1,d_0)$, $(\theta_2,d_0)$, and $(\theta_7,\md{7})$.
The translational lattice of a commensurate TBG structure is determined by its twist angle, whereas its exact point symmetry also depends on the location of the rotation axis~\cite{PhysRevB.98.085435}. 
Each unit cell was generated in the $xy$-plane from AA-stacked bilayer graphene by rotating the top layer about an axis perpendicular to the graphene sheets and passing through coincident hollow sites. 
The resulting structures have $D_6$ point-group symmetry~\cite{PhysRevB.98.085435, PhysRevB.98.245103}, allowing reciprocal-space integrations to be restricted to the corresponding irreducible wedge of the moir\'{e} Brillouin zone~(BZ). 
The layers were kept planar, and no in-plane structural relaxation was applied.
The $(\theta_7,\md{7})$ structure therefore represents a constrained flat-band model rather than an equilibrium geometry. 

\begin{figure}[htbp]
\centering
\includegraphics[width=0.99\columnwidth]{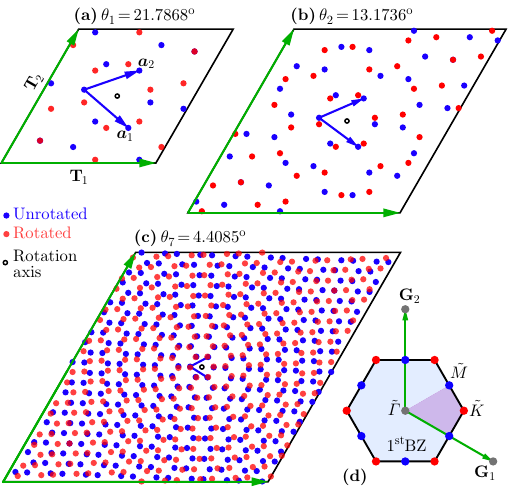}
\caption{\label{fig0}
(a)-(c) Top views of moir\'{e} unit cells for $(m,r)=(1,1)$, $(2,1)$, and $(7,1)$.
Red dots mark the rotated top layer, blue dots the unrotated bottom layer, and the open black circle marks the hollow-site rotation axis.
$\mathbf{a}_1$ and $\mathbf{a}_2$ are the bottom-layer primitive vectors.
The moir\'{e} primitive vectors $\mathbf{T}_1$ and $\mathbf{T}_2$ enclose $60^\circ$ and have length $a\sqrt{3m^2+3m+1}$, equal to (a) $6.5085$~{\AA}, (b) $10.7229$~{\AA}, and (c) $31.9800$~{\AA}, for $a=2.46$~{\AA}.
(d) Moir\'{e} reciprocal lattice with primitive vectors $\mathbf{G}_1$ and $\mathbf{G}_2$ enclosing $120^\circ$ and having magnitude $4\pi/(\sqrt{3}|\mathbf{T}_1|)$, equal to $1.1147$, $0.6766$, and $0.2269$~{\AA}$^{-1}$ for the three configurations, respectively.
The first BZ~(light blue), the irreducible wedge~(light purple), and points $\Gm$~(gray), $\Km$~(red), and $\Mm$~(blue) are indicated.
For the three configurations, $|\Gm\Mm|=|\mathbf{G}_1+\mathbf{G}_2|/2$ equals $0.5574$, $0.3383$, and $0.1134$~{\AA}$^{-1}$, while $|\Gm\Km|=|2\mathbf{G}_1+\mathbf{G}_2|/3$ equals $0.6436$, $0.3906$, and $0.1310$~{\AA}$^{-1}$, respectively.
}
\end{figure}

Figure~\ref{fig0} illustrates the three real-space lattices and their corresponding reciprocal-space construction. 

\section{Dispersive-band regime\label{sec41}}
\begin{figure}[htbp]
\centering
\includegraphics[width=0.99\columnwidth]{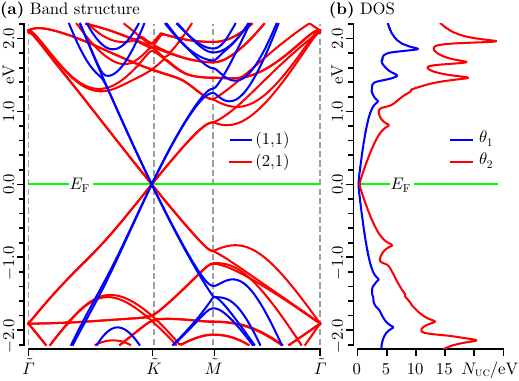}
\caption{\label{fig2A}
(a) DFT-PBE band structures for the $\theta_1$~(blue) and $\theta_2$~(red) configurations along the high-symmetry path $\Gm\Km\Mm\Gm$ of Fig.~\ref{fig0}(d).
(b) Corresponding DOS profiles, each normalized so that its integral over occupied energies equals the total number of electrons per moir\'{e} unit cell~($N_{\msc{uc}}$) and shown with a Lorentzian broadening of $0.02$~eV.
Energies in both panels are measured relative to $\EF=0$.}
\end{figure}

The electronic structure and energy-loss response are first examined for the two commensurate TBG lattices in the nonmagic, dispersive-band regime. Electronic energies are measured relative to the intrinsic Fermi energy $\EF$ of each configuration at $0$~K, determined by the charge-neutral electron count. 
At the reference separation $d_0$, moir\'{e}-induced reconstruction near $\EF$ remains sufficiently modest for the $\theta_1$ and $\theta_2$ structures to retain the characteristic semimetallic electronic structure of graphene.
This is apparent from the low-energy $\pi$ and $\pi^{*}$ band dispersions shown in Fig.~\ref{fig2A}(a), folded into the moir\'{e} BZ of Fig.~\ref{fig0}(d).
In both configurations, two nearly degenerate Dirac cones meet at $\Km$, with their vertices at $\EF$.
The underlying density of states~(DOS) vanishes at that energy, as reflected in the broadened profiles of Fig.~\ref{fig2A}(b).

To compare the Dirac dispersions independently of the different BZ sizes, average Fermi velocities of $\vF{1}=8.23\times10^5$ and $\vF{2}=8.10\times10^5$~m/s were extracted directly from the KS-band data~(see Appendix~\ref{app:eq12}).  
The resulting ${\sim}1.5\%$ decrease from $\theta_1$ to $\theta_2$ confirms that the twist-angle reduction only weakly affects the Dirac-cone dispersion in the reference-distance configurations. 
Appendix~\ref{app:eq12} shows that the directional estimates remain close to these averages, with a spread of at most approximately $4\%$ in the sampled directions, while  Fig.~\ref{fig:fermi_velocity} illustrates how the average velocities capture the central trend of the cones in both systems.
Both values also remain close to the estimate $\vFo=8.30\times10^5$~m/s obtained from DFT calculations for graphene~\cite{PhysRevB.96.201408}.

Away from $\EF$, the moir\'{e} potential induced by the relative rotation reconstructs the folded $\pi$ and $\pi^*$ manifolds through interlayer hybridization. 
Within the low-energy Dirac-derived branches, this coupling produces distinct saddle points in the band structure and the associated van Hove singularities~(VHSs) in the DOS on both the electron and hole sides.
The innermost VHS pair shifts toward $\EF$ from $\theta_1$ to $\theta_2$, narrowing the intervening low-DOS region.
At higher energies, the denser folded-band manifold of the larger $\theta_2$ supercell produces additional extrema and a denser sequence of DOS peaks.
A moderate electron-hole asymmetry is also evident from the unequal energies and intensities of the corresponding peaks.

\begin{figure}[!!h]
\centering
\includegraphics[width=0.80\columnwidth]{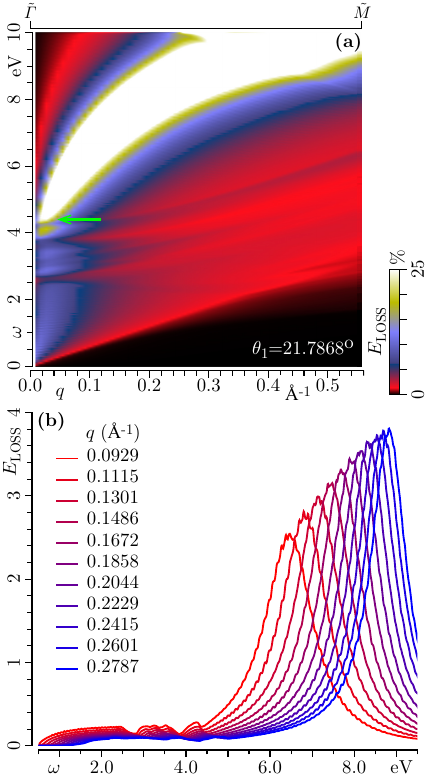}
\caption{\label{fig1} 
(a) Momentum-energy density map of $\Eloss$~[Eq.~(\ref{eqloss})] for in-plane momentum transfers $q$ along $\Gm\Mm$ and energies $\omega<10$~eV.
The upper color-scale limit is set to $25\%$ of the maximum intensity, with larger values appearing saturated.
The green arrow marks the low-$q$ $\pi$-plasmon branch; its quasi-optical limit is represented by the smallest sampled momentum, $q_{\min}=9.289\times10^{-3}$~{\AA}$^{-1}$~(see Appendix~\ref{sec3}).
(b) Unnormalized $\Eloss$ spectra at fixed momentum transfers ranging from $10\,q_{\min}$ to $30\,q_{\min}$, shown from red to blue with increasing $q$.
The line cuts highlight the progressive blue shift and intensity increase of the $\pi$-plasmon peak.}
\end{figure}

In pristine reference-distance TBG, the loss function of the $\theta_1$ configuration in Fig.~\ref{fig1} displays a dominant, strongly dispersive peak whose energy approaches ${\sim}4.4$~eV in the long-wavelength limit and undergoes a pronounced blue shift with increasing momentum transfer.

This peak is identified as the $\pi$ plasmon, whose characteristic signature in the energy-loss response of $sp^2$-bonded carbon
materials reflects a collective excitation of the $\pi$-electron
system sustained predominantly by $\pi\to\pi^\ast$ interband
transitions~\cite{PhysRev.138.A197, PhysRevLett.100.196803,
Hill_2009, PhysRevB.77.233406, PhysRevB.93.035440}.
As in monolayer and bilayer graphene, its long-wavelength energy is redshifted relative to the corresponding resonance in graphite~\cite{PhysRevB.93.035440}.
Two weaker, less dispersive features between approximately $2.6$ and $4.3$~eV are consistent with additional transition channels associated with the multiple VHSs in the TBG DOS shown in Fig.~\ref{fig2A}(b). 
At lower energies, a broad electron-hole continuum arising from single-particle excitations resembles that observed in few-layer graphene~\cite{PhysRevB.77.233406}. 

Turning to the energy-loss response of electron-doped large-angle TBG, rigid upward shifts of the Fermi level are chosen to probe both the approximately linear electron-side Dirac cones and the region beyond them.
As shown in Fig.~\ref{fig2A}, the linear dispersion extends to ${\sim}1.2$~eV above $\EF$ for $\theta_1$, but only to ${\sim}0.8$~eV for $\theta_2$.
The similar Fermi velocities nevertheless allow the low-energy electronic structures of the two reference-distance configurations to be mapped approximately onto one another. 
This correspondence arises from the geometrical relation between the real- and reciprocal-space scales of the two moir\'{e} supercells, characterized by the factor
$s^{1}_{2}=\sqrt{19/7}$, defined as the ratio of the $\theta_2$ to $\theta_1$ moir\'{e} lattice constants, or equivalently as the inverse ratio of their reciprocal-lattice scales.
Throughout the moir\'{e} BZ, the momentum coordinates of $\theta_2$ are obtained exactly from those of $\theta_1$ by division by $s^{1}_{2}$, whereas multiplication of the corresponding $\theta_2$ band and excitation energies by $s^{1}_{2}$ yields only an approximate mapping onto the $\theta_1$ energy scale, as detailed in Appendix~\ref{app:eq12}.
Outside the low-energy correspondence window, the momentum relation remains exact, whereas the energy mapping progressively deteriorates because of nonlinear and anisotropic band-structure effects.

On this basis, the shifts $\dEF{1}=0.5$, $1.0$, and $1.5$~eV are considered for $\theta_1$, placing the Fermi level well within, near the upper limit of, and beyond the linear Dirac-cone region.
These shifts introduce excess occupations of $0.22$, $0.95$, and $2.6$ electrons per moir\'{e} unit cell, in the same order.
For $\theta_2$, the scaled shifts defined by
$\dEF{2}=\dEF{1}/s^{1}_{2}$ are $0.303$, $0.607$, and $0.910$~eV, yielding excess occupations of $0.206$, $0.87$, and $2.10$ electrons per moir\'{e} unit cell. 
The comparable occupations of the first two pairs are consistent with the approximate Dirac-cone mapping, whereas the larger mismatch for the highest-doping pair signals its breakdown beyond the linear regime. 

\begin{figure*}[htbp]
\centering
\includegraphics[width=0.99\textwidth]{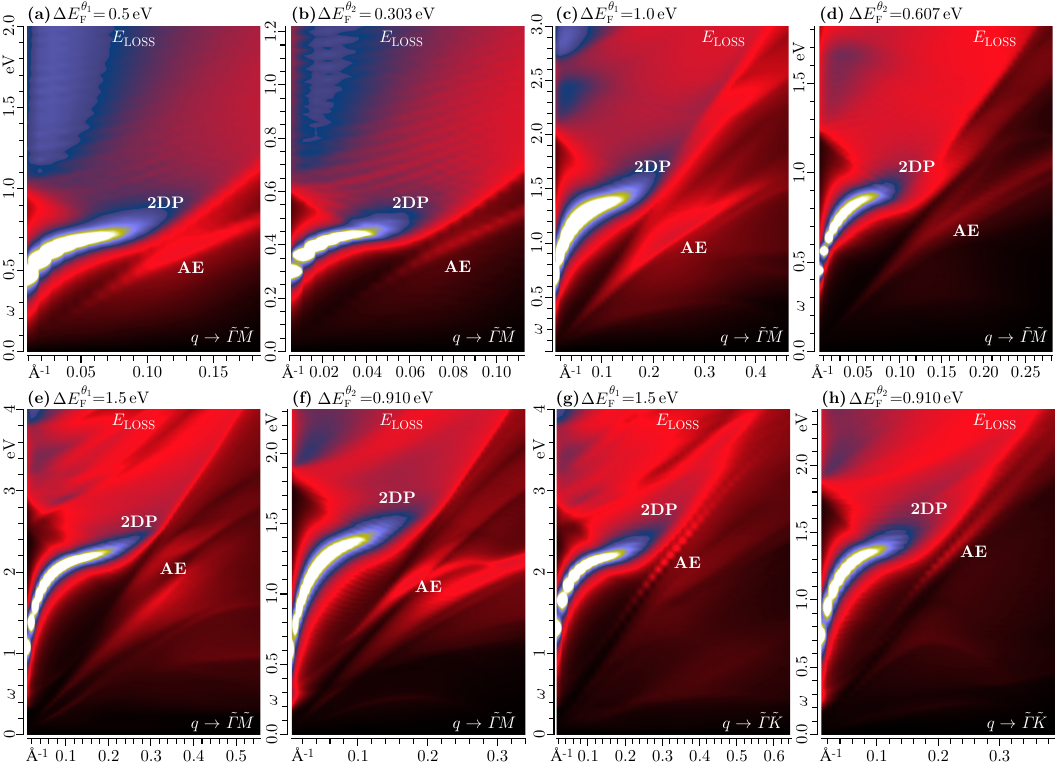}
\caption{\label{fig2c}
Momentum-energy maps of $\Eloss$ for the reference-distance $\theta_1$ and $\theta_2$ configurations at the three scaling-related doping pairs considered in the main text, defined by $\dEF{2}=\dEF{1}/s^{1}_{2}$: 
(a) $\dEF{1}=0.5$~eV and (b) $\dEF{2}=0.303$~eV; 
(c) $\dEF{1}=1.0$~eV and (d) $\dEF{2}=0.607$~eV; 
(e), (g) $\dEF{1}=1.5$~eV and (f), (h) $\dEF{2}=0.910$~eV.
Momentum transfers lie along $\Gm\Mm$ in (a)-(f) and along $\Gm\Km$ in (g) and (h). 
The corresponding high-symmetry-segment lengths, reported in Fig.~\ref{fig0}, have the common $\theta_1$-to-$\theta_2$ ratio $s^{1}_{2}$.
The 2DP and emergent AE branches are indicated.
The horizontal~($q$) and vertical~($\omega$) ranges are selected separately for each doping pair and scaled between the corresponding $\theta_1$ and $\theta_2$ panels to facilitate comparison. Each color scale is independently capped at $15\%$ of its maximum. 
All remaining computational and graphical settings follow those of Fig.~\ref{fig1}(a).}
\end{figure*}

Figure~\ref{fig2c} shows the energy-loss response for the three doping pairs, with momentum transfers along $\Gm\Mm$ in panels (a)-(f) and, for the highest-doping pair, also along $\Gm\Km$ in panels (g) and (h).
For the first two doping pairs~[Fig.~\ref{fig2c}(a)-(d)], only the $\Gm\Mm$ direction is shown because the dominant plasmonic response was verified to remain nearly isotropic while the electronic states involved retain approximately Dirac-like dispersions.
Both directions are instead shown for the third pair to expose the anisotropic response beyond the linear Dirac-cone regions.

The most prominent doping-induced feature is the emergence of a two-dimensional Dirac plasmon~(2DP), arising from the collective dynamics of charge carriers in the partially occupied Dirac bands.
Outside the intraband single-particle excitation~(SPE) continuum, its dispersion follows the characteristic $\sqrt{q}$ behavior of a two-dimensional plasmon.
Upon entering the SPE continuum, the mode undergoes Landau damping, and its broadened loss maximum bends toward an approximately linear dispersion, consistent with the behavior reported for monolayer and bilayer graphene~\cite{PhysRevB.93.035440,Pisarra_2014}.

\begin{figure}[htbp]
\centering
\includegraphics[width=0.99\columnwidth]{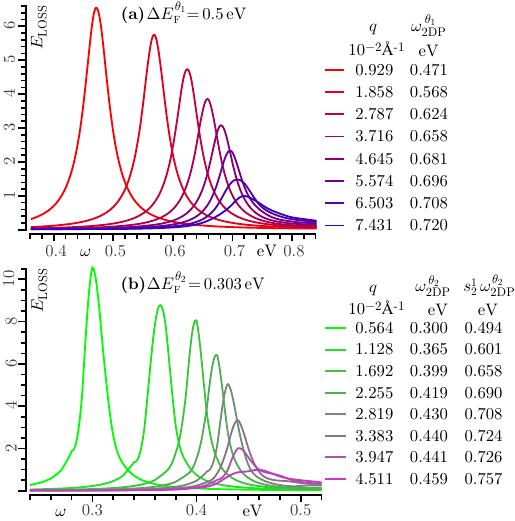}
\caption{\label{fig2d}
Electron energy-loss spectra and extracted 2DP peak energies for the lowest scaling-related doping pair: (a) $\dEF{1}=0.5$~eV and (b) $\dEF{2}=0.303$~eV.
The spectra are evaluated at corresponding momentum transfers along $\Gm\Mm$ satisfying $q_2=q_1/s^{1}_{2}$.
The momentum transfers and extracted peak energies are listed beside each panel. The final column in (b) also lists the $\theta_2$ peak energies after multiplication by $s^{1}_{2}$ for direct comparison with (a).
All remaining computational and graphical settings are as in Fig.~\ref{fig1}(b).}
\end{figure}

For the first two doping pairs~[Fig.~\ref{fig2c}(a)-(d)], the 2DP dispersions closely obey the scaling relation $\omega_{\mathrm{2DP}}^{\theta_1}(q_1)\simeq s^{1}_{2}\omega_{\mathrm{2DP}}^{\theta_2}(q_2)$ at geometrically corresponding momentum transfers satisfying   $q_1=s^{1}_{2}q_2$. 
The corresponding dispersions therefore nearly collapse onto one another when both momentum and energy are rescaled. 
This scaling is quantified in Fig.~\ref{fig2d} for the lowest doping pair using the loss spectra and extracted 2DP peak energies at geometrically corresponding momentum transfers $q_1$ and $q_2$ along $\Gm\Mm$.
Because the momentum relation is exact, this comparison directly tests the approximate scaling of the excitation energies.
The relative deviations $|\omega_{\mathrm{2DP}}^{\theta_1}(q_1)-s^{1}_{2}\omega_{\mathrm{2DP}}^{\theta_2}(q_2)|/\omega_{\mathrm{2DP}}^{\theta_1}(q_1)$ range from $2.6\%$ to $5.9\%$, with a mean value of approximately $4.6\%$.
The ratios $\omega_{\mathrm{2DP}}^{\theta_1}(q_1)/\omega_{\mathrm{2DP}}^{\theta_2}(q_2)$ nevertheless remain nearly momentum independent, averaging $1.575$, only ${\sim}4.4\%$ below $s^{1}_{2}$.
The two 2DP dispersions are therefore nearly homothetic, although their energy ratio is systematically smaller than the ideal scaling factor.

The leading scaling behavior can be understood from the long-wavelength relation
$\omega_{\mathrm{2DP}}\propto(\dEFo\,q)^{1/2}$
for an ideal two-dimensional Dirac system~\cite{PhysRevB.75.205418}.
Indeed, because the mapped configurations satisfy
$\dEF{1}=s^{1}_{2} \dEF{2}$ and
$q_1=s^{1}_{2}q_2$, the ideal relation predicts 
\begin{equation}
\frac{\omega_{\mathrm{2DP}}^{\theta_1}(q_1)}
{\omega_{\mathrm{2DP}}^{\theta_2}(q_2)}
\simeq
\left[
\frac{\dEF{1}q_1}
{\dEF{2}q_2}
\right]^{1/2}
=s^{1}_{2}.    
\end{equation}
The residual departure from this ideal result is comparable in magnitude to the few-percent spread among the directional Fermi-velocity estimates discussed in Appendix~\ref{app:eq12} and is consistent with the approximate nature of the Dirac-cone mapping. 

At the third doping pair~[Fig.~\ref{fig2c}(e)-(h)], the 2DP remains clearly identifiable and retains a qualitative scaled correspondence between the two configurations along both momentum-transfer directions. The correspondence is closest before the mode substantially overlaps the SPE continuum. 
At larger momenta, its curvature and Landau-damped continuation acquire increasingly direction- and system-dependent corrections. 
This behavior reflects the nonlinear, anisotropic, and multiband electronic structure sampled beyond the individual Dirac-cone regions.

In systems formed by two coupled layers, the conventional in-phase plasmon can be accompanied by an acoustic plasmon~(AP)~\cite{PhysRevB.23.805,PhysRevB.37.937}.
This mode originates from out-of-phase charge-density oscillations between the layers and exhibits a long-wavelength dispersion that is linear in the momentum transfer when interlayer tunneling is negligible.
Finite interlayer tunneling instead opens a long-wavelength gap in the out-of-phase branch, removing its strictly acoustic character~\cite{PhysRevLett.81.4216}.
Acoustic or acoustic-like branches can, however, also arise from multicomponent intraband dynamics without representing antisymmetric layer-density oscillations.
In extrinsic monolayer graphene, for example, an AP is generated by carriers with distinct projected Fermi velocities within the same band and disappears for momentum transfers along $\Gamma M$~\cite{Pisarra_2014}.
In doped bilayer graphene, both well-defined and damped acoustic branches have instead been associated with the simultaneous occupation of multiple bilayer-derived bands and the resulting distinct carrier components~\cite{PhysRevB.93.035440}.
For TBG, layer-resolved continuum-model calculations predict a layer-antisymmetric AP at sufficiently large twist angles, where reduced interlayer hybridization restores a well-defined layer-pseudospin character.
This mode is sustained by interlayer Coulomb coupling, while the symmetry-protected gaplessness of the Dirac cones allows the branch to remain acoustic despite finite moir\'{e} coupling. 
The predicted branch lies only slightly above the upper boundary of the SPE continuum and therefore avoids intraband Landau damping within RPA~\cite{PhysRevB.110.045431}.

By contrast, the present \emph{ab initio} loss maps display acoustic-like excitation~(AE) branches embedded within the continuum of intraband SPEs.
Their strong Landau damping, together with the absence of layer resolution in the macroscopic loss function, prevents their unambiguous identification as well-defined APs or their assignment to a specifically layer-antisymmetric origin.
The AE branches are much more sensitive than the 2DP to the momentum-transfer direction and the corresponding projected carrier velocities, as generally expected for acoustic-like excitations associated with multicomponent carrier dynamics~\cite{Pisarra_2014, PhysRevB.93.035440}.
For the lowest doping pair~[Fig.~\ref{fig2c}(a),~(b)], these branches also map closely onto one another under the same rescaling.
Starting from the second doping pair~[Fig.~\ref{fig2c}(c),~(d)], however, appreciable differences emerge because, after rescaling, this pair lies beyond the empirical correspondence window identified in Appendix~\ref{app:eq12}, even though both Fermi levels remain within their respective Dirac-cone regions.
These differences mark the onset of deterioration in the low-energy electronic correspondence before the individual band dispersions depart completely from their approximately conical form. 
At the third doping pair~[Fig.~\ref{fig2c}(e)-(h)], the AE correspondence becomes strongly direction dependent. Along $\Gm\Km$~[Fig.~\ref{fig2c}(g),~(h)], the leading AE ridge retains a close correspondence under the momentum-energy rescaling, consistent with similar projected carrier velocities in this direction. Along $\Gm\Mm$~[Fig.~\ref{fig2c}(e),~(f)], by contrast, several approximately linear loss ridges are resolved, particularly clearly for $\theta_2$. 
Their multiplicity is consistent with the additional intraband transition channels generated by partially occupied folded bands with distinct projected carrier velocities beyond the Dirac-cone region. 
The differences in the resulting spectral patterns and slopes provide a pronounced manifestation of the breakdown of the low-energy electronic correspondence.
Taken together, the three doping pairs trace its progressive deterioration from an approximately isotropic Dirac response to anisotropic, configuration-specific multiband dynamics.

\section{Magic-distance configuration\label{sec42}}
The flat-band regime is examined using the $(\theta_7,\md{7})\equiv(4.4085^\circ,2.587~\text{\AA})$ configuration, whose $676$-atom moir\'{e} cell provides a computationally tractable alternative to the commensurate $(31,1)$ cell at $\theta_{31}=1.050^\circ$, representative of magic-angle TBG at the reference interlayer distance $d_0$~\cite{Palamara2025}.
The latter contains $11,908$ atoms and therefore remains beyond the practical reach of a direct PW-TDDFT-RPA treatment.
\begin{figure}[htbp]
\centering
\includegraphics[width=0.99\columnwidth]{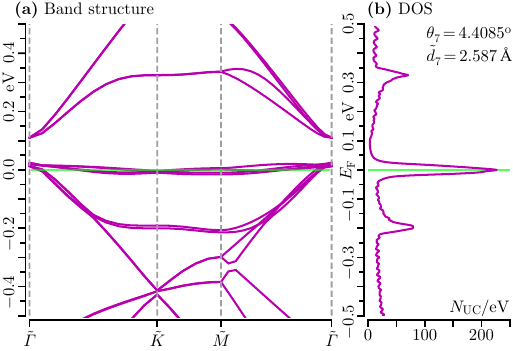}
\caption{\label{fig4}
(a) DFT-PBE band structure for the
$(\theta_7,\md{7})=(4.4085^\circ,2.587~\text{\AA})$
configuration along the high-symmetry path
$\Gm\Km\Mm\Gm$ of Fig.~\ref{fig0}(d).
(b) Corresponding DOS profile, normalized so that its integral over
occupied energies equals the total number of electrons per moir\'{e}
unit cell~($N_{\msc{uc}}$).
A Lorentzian broadening of $0.005$~eV, reduced relative to that used
in Fig.~\ref{fig2A}, is adopted to resolve the fine structure
associated with the four quasi-flat bands.
Energies in both panels are measured relative to $\EF=0$.}
\end{figure}

Figure~\ref{fig4} shows the DFT-PBE band structure and the corresponding DOS of the $(\theta_7,\md{7})$ configuration.
At the magic distance, the four bands form a narrow quasi-flat
manifold around the intrinsic Fermi energy, $\EF=0$, yielding  
a pronounced DOS peak.
A comparable low-energy enhancement persists over a finite interval of interlayer distances around $\md{7}$, reflecting the broad minimum of the quasi-flat-band width reported for this configuration~\cite{Palamara2025}.
Additional DOS maxima occur near $+0.3$~eV and $-0.2$~eV; unlike the central peak, these satellite features vary appreciably with the interlayer distance in the vicinity of $\md{7}$.

This behavior differs qualitatively from that of the reference-distance configurations shown in Fig.~\ref{fig2A}, where the Dirac crossing is accompanied by a vanishing DOS at $\EF$.
Furthermore, scaling the band energies relative to $\EF$ by the factor
$s^{31}_7=0.2383$, defined within the Bistritzer-MacDonald~(BM)
continuum model~\cite{doi:10.1073/pnas.1108174108}, provides an
approximate mapping of the quasi-flat manifold onto that of the
reference-distance magic-angle configuration~\cite{Palamara2025}.
\begin{figure}[h!]
\centering
\includegraphics[width=0.99\columnwidth]{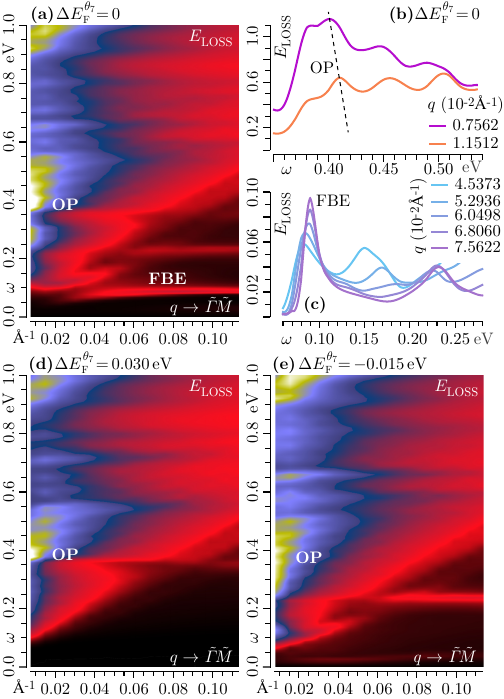}
\caption{\label{fig5}
Momentum-energy maps of $\Eloss$ for the
$(\theta_7,\md{7})$ configuration at
(a) charge neutrality, $\dEF{7}=0$, and under rigid Fermi-level
shifts of (d) $\dEF{7}=0.03$~eV and
(e) $\dEF{7}=-0.015$~eV.
Fixed-$q$ loss profiles for the intrinsic configuration in the
energy ranges of the (b) OP and (c) FBE features.
Momentum transfers lie along $\Gm\Mm$.
The OP and FBE features are indicated.
Each color scale is independently capped at $25\%$ of its maximum.
All remaining computational and graphical settings, including the
Lorentzian broadening, follow those of
Figs.~\ref{fig1}(a) and~\ref{fig2c}.}
\end{figure}

The low-energy electron energy-loss function of the magic-distance configuration is next examined at charge neutrality and for two rigid Fermi-level shifts, $\dEF{7}=0.03$~eV and $-0.015$~eV.
With spin degeneracy included, these shifts correspond, respectively, to the addition of $2.000$ and the removal of $2.098$ electron pairs per moir\'{e} cell.
The resulting loss maps and selected fixed-$q$ profiles are shown in Fig.~\ref{fig5}.
At intrinsic filling, the Fermi level intersects the four quasi-flat bands, suggesting that the partially occupied manifold could support a low-energy collective charge excitation.
A low-energy loss feature is indeed observed in Figs.~\ref{fig5}(a) and~\ref{fig5}(c) and exhibits qualitative characteristics reminiscent of the narrow-band plasmon predicted in Ref.~\cite{doi:10.1073/pnas.1909069116}. 
Its intensity tends to vanish in the long-wavelength limit, whereas at larger $q$ it gains spectral weight and becomes nearly nondispersive.
However, as shown by the dielectric analysis in Appendix~\ref{app:dielectric}, $\re[\epsilon_{\mathrm M}(q,\omega)]$ exhibits a marked variation across the FBE energy range but no clear zero crossing at the corresponding loss maxima, while $\im[\epsilon_{\mathrm M}(q,\omega)]$ remains appreciable. The dielectric response therefore does not provide an unambiguous signature of a well-defined plasmon, although some collective contribution cannot be excluded. The feature is consequently denoted a flat-band excitation~(FBE) and interpreted in terms of a combination of intra- and interband transitions involving the quasi-flat manifold, with nearby high-DOS regions enhancing the interband contribution. 

By contrast, a distinct finite-energy loss peak is resolved in Fig.~\ref{fig5}(b) at $\omega_{\mathrm{OP}}\simeq0.40$~eV for the smallest momentum transfer accessible on the BZ mesh~($0.00756$~{\AA}$^{-1}$ along $\Gm\Mm$, discussed in Appendix~\ref{sec3}). 
As shown in Appendix~\ref{app:dielectric}, $\re[\epsilon_{\mathrm M}(q,\omega)]$ changes sign across the OP energy range, and the loss maximum lies near its higher-energy zero crossing, where $\im[\epsilon_{\mathrm M}(q,\omega)]$ has decreased from the neighboring absorption maximum. 

This dielectric signature supports the identification of the mode as an interband plasmon, here denoted the optical plasmon~(OP).
The OP is attributed primarily to interband transitions connecting the quasi-flat manifold with neighboring high-DOS regions.
As a heuristic energy-scale comparison, applying the band-energy factor $s^{31}_7=0.2383$ to $\omega_{\mathrm{OP}}$ gives $\tilde{\omega}_{\mathrm{OP}}\simeq0.095$~eV, which lies below the $0.18$-$0.23$~eV interband collective excitations observed in TBG at $\theta=1.35^\circ$~\cite{Hesp2021}. 
The difference is plausible because the experimental twist angle exceeds that of the $(\theta_{31},d_0)$ configuration onto which the magic-distance OP energy is mapped, and a single band-energy factor cannot account for the associated changes in band reconstruction and dielectric screening.

When the quasi-flat bands are completely filled at $\dEF{7}=0.03$~eV, the FBE disappears~[Fig.~\ref{fig5}(d)].
Its suppression supports its association with the partially occupied quasi-flat manifold. 
The OP remains clearly visible because interband transitions from the occupied quasi-flat bands to the unoccupied high-DOS states near $0.3$~eV remain available.

For $\dEF{7}=-0.015$~eV, the rigid shift depletes the quasi-flat manifold and also depopulates the uppermost $0.015$-eV portion of the three dispersive valence bands that cross the shifted Fermi level near $\Gm$ and touch the quasi-flat bands~[Fig.~\ref{fig4}(a)].
Of the $2.098$ electron pairs removed in total, $2.000$ account for
depletion of the quasi-flat manifold, whereas the remaining
$0.098$ pair is removed from these dispersive states. This small
population of dispersive-band holes produces the faint low-energy
branch visible at small $q$ in Fig.~\ref{fig5}(e). 
The OP also persists but extends over a broader energy range, reflecting the larger set of available interband transitions from the occupied high-DOS states near $-0.2$~eV into the depleted quasi-flat manifold.

The filling dependence therefore distinguishes the transition-driven FBE from the more robust interband OP and reveals how occupation of the quasi-flat manifold redistributes low-energy spectral weight.

\section{Conclusions\label{sec5}}
The dielectric response and plasmonic excitations were investigated in representative TBG realizations of the dispersive- and flat-band regimes. 

Reference-distance TBG configurations reveal a hierarchy between robust graphene-derived charge dynamics and moir\'{e}-specific corrections.
The $\pi$ plasmon retains its graphene-like character despite its redshift relative to graphite, whereas the effects of moir\'{e}-induced band folding and hybridization are manifested primarily in weaker loss features associated with additional VHSs. 
Upon electron doping, the near-isotropy and approximate two-angle scaling of the 2DP within the linear Dirac regime show that the leading low-energy response is governed mainly by Dirac-cone kinematics: the rescaling of momentum and Fermi-level shift, together with the nearly equal Fermi velocities, accounts for the correspondence between the plasmon resonances. 
The progressive breakdown of this correspondence at higher doping levels signals the growing influence of nonlinear, anisotropic, and multiband corrections.
The AE branches provide a more sensitive probe of these corrections: their strong Landau damping and directional dependence are consistent with multicomponent intraband dynamics and prevent their unambiguous identification as well-defined layer-antisymmetric acoustic plasmons. 

The magic-distance response is instead organized by the quasi-flat manifold and the neighboring high-DOS states.
The FBE exhibits qualitative similarities to the narrow-band plasmon predicted within a continuum-model RPA treatment~\cite{doi:10.1073/pnas.1909069116}, but its dielectric signature in the present atomistic calculation does not establish a well-defined plasmon. This difference suggests sensitivity to the detailed band structure and available electron-hole transition channels, while also reflecting the distinct geometry and filling considered here. 
The OP, on the other hand, is accompanied by a dielectric zero and persists across all three filling conditions examined because the relevant interband transition channels remain available.
Changes in filling therefore shift its energy and redistribute its spectral weight without suppressing the mode.
Rescaling the characteristic OP energy to the reference-distance $\theta_{31}\simeq1.05^\circ$ configuration places it in the mid-infrared, below the interband collective excitations measured at the larger angle $\theta\simeq1.35^\circ$~\cite{Hesp2021}. 
This ordering is consistent with the anticipated reduction of the characteristic resonance energy with decreasing twist angle. 
A closer comparison could be pursued by extending the distance-based mapping to the experimental angle, closely represented by $\theta_{24}=1.3501^{\circ}$, using a different reduced interlayer separation at $\theta_7$ and recalculating the corresponding dielectric response.

These results have two broader implications.
First, combining the magic-distance construction established in Ref.~\cite{Palamara2025} with twist-angle scaling offers a computationally tractable route for estimating selected plasmonic energy scales of magic-angle TBG at the reference interlayer separation.
Second, the magic-distance results suggest that the interlayer separation may provide an additional control parameter for shifting collective excitations toward magic-angle energy scales, potentially easing the need for precise twist-angle control.
The experimental feasibility of the required compression and the effects of the accompanying structural relaxation must nevertheless be established before this possibility can be assessed quantitatively.

Finally, the \textit{ab initio} dynamical screening obtained here provides a starting point for assessing a possible plasmon-mediated contribution to Cooper pairing in magic-angle TBG~\cite{PhysRevResearch.2.022040,PhysRevB.103.235401,PhysRevB.109.045404}.
Incorporating the resulting frequency-dependent screened interaction into superconducting density-functional theory~\cite{PhysRevLett.60.2430,PhysRevLett.111.057006,PhysRevB.102.214508} could determine whether the collective modes identified here enhance or suppress pairing and clarify their interplay with phonon-mediated interactions.
\section*{Acknowledgments}
The research was partially supported by the \emph{Centro Nazionale di Ricerca in High-Performance Computing, Big Data and Quantum Computing}, PNRR 4 2 1.4, CI CN00000013, CUP H23C22000360005.
The authors acknowledge the \emph{Marconi}, \emph{Marconi}100, \emph{Galileo}100, and \textit{Leonardo} high-performance computing resources, provided by the \href{\cinecaurl}{CINECA consortium (Italy)}, within the {\infnprog} project, under the \href{\cinecainfn}{CINECA-INFN} agreement.
The authors also acknowledge the \emph{Newton} and \emph{Alarico} high-performance computing clusters, provided by the University of Calabria.
\bibliography{Refs}
\begin{appendix}
\section{Ground-state calculations\label{sec2b}}
The electronic ground states of the $(\theta_1,d_0)$, $(\theta_2,d_0)$, and $(\theta_7,\md{7})$ configurations   were calculated within a PW-DFT framework using the PBE exchange-correlation functional~\cite{PhysRevLett.77.3865}, as implemented in the Quantum ESPRESSO suite~\cite{giannozzi2009quantum, giannozzi2017advanced}. 
Optimized norm-conserving Vanderbilt pseudopotentials were employed to describe the electron-ion interaction~\cite{PhysRevB.88.085117}. 
The periodic replicas of each TBG slab were separated by $20$~{\AA} of vacuum along the out-of-plane $z$ direction.
Kinetic-energy cutoffs of $80$~Ry for the KS wave functions and $640$~Ry for the charge density were used for all three configurations.  
Marzari-Vanderbilt cold smearing with a width of $0.01$~Ry was used to facilitate self-consistent-field~(SCF) convergence, with a convergence threshold of $10^{-6}$~Ry on the estimated energy error. 
BZ integrations were performed using Monkhorst-Pack grids~\cite{monkhorst1976special} of $18\times18\times1$ points for the reference-distance $\theta_1$ and $\theta_2$ configurations and $12\times12\times1$ points for the magic-distance $\theta_7$ configuration.
The $D_6$ point-group symmetry was used to restrict BZ sampling to the integration wedge shown in Fig.~\ref{fig0}. 

Following the SCF calculations, non-self-consistent-field~(NSCF) calculations were performed on denser uniform $\mathbf{k}$-point meshes to obtain the KS eigenvalues $\varepsilon_{\nu\mathbf{k}}$ and eigenfunctions $\psi_{\nu\mathbf{k}}$ required for the dielectric-response calculations. 
Here, $\mathbf{k}$ denotes the Bloch wave vector in the first BZ, while $\nu$ labels the occupied and unoccupied KS bands. 
For $(\theta_1,d_0)$ and $(\theta_2,d_0)$, a $120\times120\times1$ mesh was used, retaining $150$ and $175$ bands, respectively. 
For $(\theta_7,\md{7})$, a $30\times30\times1$ mesh was used with $1800$ bands.  
The resulting band structures and DOS profiles are analyzed together with the momentum-resolved loss spectra in Secs.~\ref{sec41} and~\ref{sec42}.

The computational and storage demands of the $676$-atom $(\theta_7,\md{7})$ configuration necessitated a smaller number of $\mathbf{k}$ points, although its smaller moir\'{e} BZ keeps the absolute $\mathbf{k}$-point spacing comparable to that used for the two large-angle structures.
At the $80$-Ry cutoff, storing all $1800$ KS wave functions $\{\psi_{\nu\mathbf{k}}\}_{\nu=1}^{1800}$ at a single $\mathbf{k}$ point requires approximately $42.5$~GB.
To reduce this storage requirement, their PW representation was truncated to an auxiliary kinetic-energy cutoff of $30$~Ry for the response calculation, preserving the norm of each wave function to within ${\sim}10^{-4}$.
This truncation was applied only after the NSCF calculation. 
Both the SCF and NSCF calculations retained the $80$-Ry cutoff specified above.

\section{Dynamical dielectric response\label{sec3}}
The macroscopic dielectric function and the corresponding loss function introduced in Eq.~(\ref{eqloss}) were calculated within TDDFT-RPA using an in-house computational framework developed in earlier studies~\cite{Pisarra_2014,PhysRevLett.117.116801,PhysRevB.93.035440,PhysRevB.96.201408,PhysRevB.100.235422} and adapted to handle the large $(\theta_7,\md{7})$ supercell. 

The starting point of this approach is the density-density response function $\chi^{\mathrm{KS}}$ of the auxiliary noninteracting KS system. 
In reciprocal space and Hartree atomic units, this quantity is given by the Adler-Wiser expression~\cite{PhysRev.126.413,PhysRev.129.62}:
\begin{align}
\label{eq:chiKS}
\chi^{\mathrm{KS}}_{\mathbf{G}\mathbf{G}^{\prime}}(\mathbf{q},\omega)={}&
\frac{2}{\Omega}\sum_{\mathbf{k}\in\mathrm{BZ}} \sum_{\nu,\nu^{\prime}}
\rho^{\mathbf{kq}}_{\nu\nu^{\prime}}(\mathbf{G})\,
\rho^{\mathbf{kq}}_{\nu\nu^{\prime}}(\mathbf{G}^{\prime})^*
\\&\qquad\qquad\qquad\times
\frac{
f_{\nu\mathbf{k}}-f_{\nu^{\prime}\mathbf{k}+\mathbf{q}}
}{
\omega +
\varepsilon_{\nu\mathbf{k}} -
\varepsilon_{\nu^{\prime}\mathbf{k}+\mathbf{q}} +
\mathrm{i}\,\eta}
.\notag
\end{align}
Here, $\mathbf{q}$ and $\omega$ denote the momentum and energy transferred to the system, respectively, with $\mathbf{q}$ reduced to the first BZ. $\mathbf{G}$ and $\mathbf{G}^{\prime}$ are reciprocal-lattice vectors.
The factor of $2$ accounts for spin degeneracy. 
$\Omega=N_{\mathrm{BZ}}\Omega_0$ denotes the Born–von K\'{a}rm\'{a}n normalization volume, where $\Omega_0$ represents the TBG supercell volume and $N_{\mathrm{BZ}}$ the total number of points in the uniform MP mesh covering the full BZ. 
The $\mathbf{k}$-point sum runs over this complete mesh, while both band indices, $\nu$ and $\nu^{\prime}$, span all KS bands retained in the NSCF calculations. 
The transition matrix elements between the KS states take the form:
\begin{equation}
\rho^{\mathbf{kq}}_{\nu\nu^{\prime}}(\mathbf{G})=
\left\langle
\psi_{\nu\mathbf{k}}
\right|
e^{-\mathrm{i}(\mathbf{q}+\mathbf{G})\cdot\mathbf{r}}
\left|
\psi_{\nu^{\prime}\mathbf{k}+\mathbf{q}}
\right\rangle,
\label{eq:transition_density}
\end{equation}
where $\mathbf{q}+\mathbf{G}$ represents the momentum carried by a microscopic Fourier component of the perturbation. 
The occupation factors were evaluated using the Fermi-Dirac distribution at $T=10^{-4}$~K, effectively corresponding to the zero-temperature limit, independently of the cold smearing used in the ground-state calculations. 
Electron and hole doping were modeled within the rigid-band approximation by shifting the chemical potential by $\dEFo$ relative to its charge-neutral value in the occupation factors entering Eq.~(\ref{eq:chiKS}).
Self-consistent doping-induced changes in the KS potential, band dispersions, charge density, and atomic structure were therefore not included.
The parameter $\eta>0$ introduces Lorentzian spectral broadening, with $\eta=0.02$~eV used throughout. 
The formal retarded-response limit is recovered as $\eta\rightarrow0^+$.  
In practice, the KS states on the full $\mathbf{k}$-point mesh were reconstructed from those in the integration wedge shown in Fig.~\ref{fig0} using the crystal symmetries of the commensurate structure. The corresponding energies, occupations, and transition matrix elements were then used to evaluate Eq.~(\ref{eq:chiKS}). Each point of the full-BZ mesh enters the sum once, without additional symmetry weights. 

While Eq.~(\ref{eq:chiKS}) describes the independent-particle response of the KS electrons, the interacting density-density response function $\chi$ is obtained from the TDDFT Dyson equation involving $\chi^{\mathrm{KS}}$ and the Hartree-exchange-correlation kernel~\cite{PhysRevLett.76.1212}. 
Within the RPA, the exchange-correlation contribution to the kernel is neglected, while the Hartree contribution is retained, yielding
\begin{align}
\label{chiint}
\chi_{\mathbf{G}\mathbf{G}^{\prime}}
(\mathbf{q},\omega){}&=
\chi^{\mathrm{KS}}_{\mathbf{G}\mathbf{G}^{\prime}}
(\mathbf{q},\omega)
\\&
+\sum_{\mathbf{G}_1,\mathbf{G}_2}
\chi^{\mathrm{KS}}_{\mathbf{G}\mathbf{G}_
1}
(\mathbf{q},\omega)
v_{\mathbf{G}_1\mathbf{G}_2}(\mathbf{q})
\chi_{\mathbf{G}_2\mathbf{G}^{\prime}}
(\mathbf{q},\omega).\notag
\end{align}
For a fully periodic three-dimensional system, the bare Coulomb interaction is represented by the diagonal reciprocal-space matrix elements
\begin{equation}
\label{eq:coulomb_3D}
v_{\mathbf{G}_1\mathbf{G}_2}(\mathbf{q})=
\frac{4\pi\delta_{\mathbf{G}_1\mathbf{G}_2}
}
{\left|\mathbf{q}+\mathbf{G}_1\right|^2}.
\end{equation} 
Its long-range character couples the induced densities of neighboring TBG slabs even when their electronic wave functions have negligible overlap at the chosen vacuum separation.
This spurious interaction between periodically repeated images decreases slowly with increasing vacuum thickness, making convergence toward the isolated-slab limit computationally demanding, particularly at small momentum transfers.

To describe an isolated TBG slab, the Coulomb kernel in Eq.~(\ref{eq:coulomb_3D}) was therefore replaced by a slab-truncated kernel, constructed in a mixed in-plane reciprocal-space and out-of-plane real-space representation~\cite{PhysRevB.93.035440,PhysRevB.86.165419,PhysRevB.87.075447,PhysRevB.91.195407,PhysRevB.96.201408,PhysRevB.100.235422}.
For an in-plane momentum transfer $\mathbf{q}=(q_x,q_y,0)$, the reciprocal-lattice vectors are decomposed as $\mathbf{G}_i=\mathbf{g}_i+(0,0,G_{iz})$ for $i=1,2$, where $\mathbf{g}_i=(G_{ix},G_{iy},0)$ represent the in-plane projection and $G_{iz}$ the out-of-plane component, respectively. 
In the PW representation, the slab-truncated Coulomb matrix is then
\begin{align}
\label{eq:coulomb_truncated}
v^{\mathrm{2D}}_{\mathbf{G}_1\mathbf{G}_2}(\mathbf{q})={}&
\frac{1}{L_z}
\int_{-L_z/2}^{L_z/2} dz_1\,
\int_{-L_z/2}^{L_z/2} dz_2\,
e^{-\mathrm{i}G_{1z}z_1}
\\&\qquad\qquad\qquad\quad\times
\overline{v}_{\mathbf{g}_1\mathbf{g}_2}(\mathbf{q};z_1,z_2)
e^{\mathrm{i}G_{2z}z_2},\notag
\end{align}
where $L_z$ denotes the out-of-plane length of the supercell and
\begin{equation}
\overline{v}_{\mathbf{g}_1\mathbf{g}_2}
(\mathbf{q};z_1,z_2)=
\frac{
2\pi\delta_{\mathbf{g}_1\mathbf{g}_2}
}{
\left|\mathbf{q}+\mathbf{g}_1\right|
}
e^{-\left|\mathbf{q}+\mathbf{g}_1\right|
\left|z_1-z_2\right|}
\label{eq:coulomb_mixed}
\end{equation}
is the in-plane Fourier transform of the bare Coulomb interaction. 
Because the mixed-representation kernel is nonperiodic along $z$, restricting both out-of-plane coordinates to the supercell containing the TBG slab excludes Coulomb coupling between induced densities in different periodic images.
The interacting response of the isolated slab is consequently obtained from Eq.~(\ref{chiint}) by replacing $v_{\mathbf{G}_1\mathbf{G}_2}$ with $v^{\mathrm{2D}}_{\mathbf{G}_1\mathbf{G}_2}$.

Within this formulation, the microscopic RPA dielectric matrix is
\begin{equation}
\epsilon^{\mathrm{RPA}}_{\mathbf{G}\mathbf{G}^{\prime}}
(\mathbf{q},\omega)=
\delta_{\mathbf{G}\mathbf{G}^{\prime}}-
\sum_{\mathbf{G}_1}
v^{\mathrm{2D}}_{\mathbf{G}\mathbf{G}_1}(\mathbf{q})
\chi^{\mathrm{KS}}_{\mathbf{G}_1\mathbf{G}^{\prime}}
(\mathbf{q},\omega).
\label{eq:dielectric_matrix}
\end{equation}
and its inverse satisfies 
\begin{align}
\label{eq:inverse_dielectric}
\left[
\epsilon^{\mathrm{RPA}}(\mathbf{q},\omega)^{-1}
\right]_{\mathbf{G}\mathbf{G}^{\prime}}=
\delta_{\mathbf{G}\mathbf{G}^{\prime}}
+\sum_{\mathbf{G}_1}
v^{\mathrm{2D}}_{\mathbf{G}\mathbf{G}_1}(\mathbf{q})
\chi_{\mathbf{G}_1\mathbf{G}^{\prime}}
(\mathbf{q},\omega).
\end{align}
The corresponding macroscopic dielectric function is defined as
\begin{equation}
\epsM(\mathbf{q},\omega)=
\frac{1}{
\left[
\epsilon^{\mathrm{RPA}}(\mathbf{q},\omega)^{-1}
\right]_{\mathbf{0}\mathbf{0}}
},
\label{eq:macroscopic_dielectric}
\end{equation}
The dielectric matrix is inverted before its $\mathbf{G}=\mathbf{G}^{\prime}=\mathbf{0}$ component is extracted, retaining the coupling between microscopic Fourier components and thereby incorporating crystal local-field effects~\cite{PhysRevLett.100.196803}.
The loss function is subsequently obtained from Eq.~(\ref{eqloss}).

For these calculations, the interacting response of Eq.~(\ref{chiint}) was represented in a finite reciprocal-space basis containing approximately $67$ vectors for the $\theta_1$ and $\theta_2$ configurations and $33$ vectors for the $\theta_7$ configuration.
The vectors were selected in order of increasing magnitude, retaining complete shells.
All available momentum transfers compatible with the sampled MP meshes were considered along the $\Gm\Mm$ and $\Gm\Km$ directions.
The energy transfer $\omega$ was sampled in increments of $0.01$~eV up to $10$~eV for the large-angle configurations and $4$~eV for the small-angle configuration.
For the large-angle configurations, basis-convergence tests were performed at selected $\mathbf{q}$ values using coarser energy grids.
Increasing the basis to $100$ reciprocal-lattice vectors changed both $\epsM$ and $\Eloss$ by less than $0.01\%$ at the tested points.

\section{Geometric scaling and Dirac-band correspondence in large-angle TBG configurations\label{app:eq12}}
The characteristic double-Dirac-cone structures shown in Fig.~\ref{fig2A} identify a low-energy region in which the single-particle electronic structures of the commensurate $(m,r)=(1,1)$ and $(2,1)$ TBG systems are approximately related by a joint momentum and energy scaling.
The geometrical factor underlying this construction, $s^{1}_{2}=\sqrt{19/7}=1.64751$, is the exact ratio of the reciprocal-space primitive-vector magnitudes and the inverse ratio of the real-space primitive-vector lengths of the two commensurate supercells, as discussed in section \ref{sec2a} and reported in Fig.~\ref{fig0}.
Consequently, the lengths of the high-symmetry segments $\Gm\Mm$, $\Gm\Km$, and $\Km\Mm$ scale by the same factor.

Within the BM continuum model~\cite{doi:10.1073/pnas.1108174108} and the associated equivalence mappings extended to atomistic and DFT descriptions~\cite{Palamara2025}, this factor is equivalently expressed as the ratio of the Dirac-point separations between the layers in the two configurations:
$s^{1}_{2}=\sin(\theta_1/2)/\sin(\theta_2/2)$.
In the present case, however, both configurations are maintained at the same reference interlayer distance.
Their electronic structures therefore do not belong to the same equivalence class because the interlayer distance, which controls the interlayer coupling, is held fixed rather than adjusted to compensate for the change in twist angle.
The energy mapping relies instead on the similarity of their low-energy Dirac dispersions.
\begin{figure}[htbp]
\centering
\includegraphics[width=0.99\columnwidth]{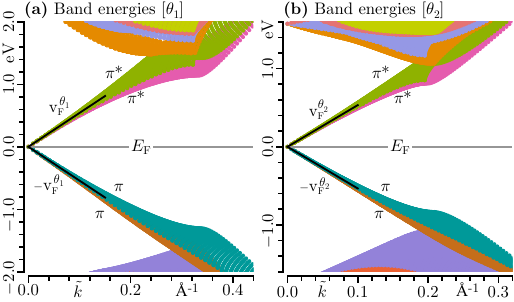}
\caption{\label{fig:fermi_velocity}
KS energies for the (a) $\theta_1$ and (b) $\theta_2$ reference-distance configurations, plotted relative to $\EF=0$ as functions of the wave-vector displacement magnitude $\tilde{k}=|\mathbf{k}-\tilde{\mathbf{K}}|$, for all directions represented in the NSCF meshes of the two systems~(see Appendix \ref{sec2b}). Away from $\Km$, the low-energy manifold resolves into two occupied~$\pi$ and two unoccupied~$\pi^*$ branches forming nearly degenerate Dirac cones.
The two black lines in each panel represent the linearized Dirac dispersions constructed from the corresponding combined Fermi-velocity estimate. 
The resulting slope magnitudes are $\hbar\vF{1}=5.4149$~eV\,{\AA} and $\hbar\vF{2}=5.3332$~eV\,{\AA}.
For both configurations, the largest and smallest directional velocities occur along $\Km\Gm$ and $\Km\Mm$, respectively.}
\end{figure}

This similarity is quantified by Fermi velocities extracted from linear fits to the five sampled points closest to $\Km$.
For the $\theta_1$ and $\theta_2$ configurations, respectively, the fits yield $8.375\times10^5$ and $8.150\times10^5$~m/s along $\Km\Gm$, and $8.078\times10^5$ and $8.054\times10^5$~m/s along $\Km\Mm$.
Along the direction through $\Km$ parallel to $\Gm\Mm$, the corresponding values are $8.227\times10^5$ and $8.105\times10^5$~m/s.
Combining the three directional fits gives $\vF{1}=8.2267\times10^5$ and $\vF{2}=8.1026\times10^5$~m/s, as also reported in section \ref{sec41}.
The $\Km$-centered dispersions in Fig.~\ref{fig:fermi_velocity} show that the combined velocities represent the mean Dirac-cone slopes over the full angular sampling. 

\begin{figure}[htbp]
\centering
\includegraphics[width=0.99\columnwidth]{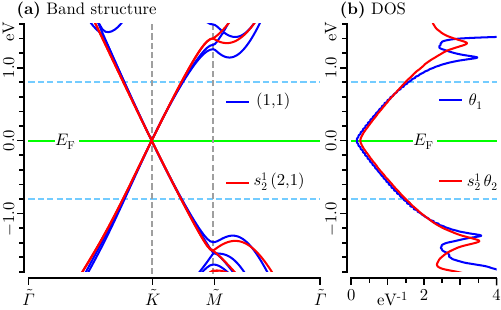}
\caption{\label{fig2B}
(a) DFT-PBE band structures and (b) DOS for $\theta_1$~(blue; cf. Fig.~\ref{fig2A}) and the rescaled $\theta_2$~(red).
For the latter, the momentum and energy coordinates are multiplied by $s^{1}_{2}$, while the DOS amplitude is divided by the same factor to preserve its integrated weight.
The green solid line marks the intrinsic Fermi level $\EF$, and the cyan dashed lines delimit the empirical correspondence window $|E-\EF|\lesssim0.8$~eV.}
\end{figure}

The near equality of $\vF{1}$ and $\vF{2}$ therefore supports an approximate energy mapping between corresponding Dirac-derived bands $\varepsilon_{\nu\mathbf{k}}^{\theta_1}$ and $\varepsilon_{\nu\mathbf{k}}^{\theta_2}$, measured relative to $\EF$, with $\nu$ labeling the two $\pi$ and two $\pi^\ast$ branches.
This relation takes the form $\varepsilon_{\nu\mathbf{k}_1}^{\theta_1}\simeq s^{1}_{2}\varepsilon_{\nu\mathbf{k}_2}^{\theta_2}$ for moir\'{e} BZ coordinates related by $\mathbf{k}_1=s^{1}_{2}\mathbf{k}_2$.
The mapping is demonstrated in Fig.~\ref{fig2B}, where the momentum and energy coordinates of the $\theta_2$ band structure are multiplied by $s^{1}_{2}$ and thereby expressed on the corresponding $\theta_1$ scales. 
Under the same energy transformation, the mapped $\theta_2$ DOS amplitude is divided by $s^{1}_{2}$, thereby preserving its integrated weight.

On their native energy scales, the approximately linear electron-side Dirac-cone regions extend to about $1.2$~eV above $\EF$ for $\theta_1$ and $0.8$~eV for $\theta_2$, as apparent from Fig.~\ref{fig2A}(a). 
Because the $\theta_2$ energy coordinate is multiplied by $s^{1}_{2}$ in the mapping, its native $0.8$~eV limit corresponds to approximately $1.32$~eV on the $\theta_1$ scale.
The geometrical scaling and ideal linear-cone approximation would therefore permit a common electron-side mapped range extending to approximately $1.2$~eV.
In the DFT results, however, the rescaled band structures and DOS profiles begin to exhibit visible system-specific differences outside the more conservative interval $|E-\EF|\lesssim0.8$~eV on the mapped $\theta_1$ energy scale.
This interval is consequently adopted as the empirical validity range of the approximate electronic correspondence, rather than as a cutoff imposed directly by the geometrical transformation. 
Beyond it, nonlinearities, anisotropies, and system-specific folded-band hybridizations progressively degrade the mapping even while the individual dispersions remain broadly cone-like.

\section{Low-energy excitations in the magic-distance configuration\label{app:dielectric}}
The nature of the low-energy features identified in Fig.~\ref{fig5} is further assessed by analyzing the real and imaginary parts of the macroscopic dielectric function $\epsilon_{\mathrm M}(q,\omega)$ of the intrinsic magic-distance configuration for momentum transfers $q\parallel\Gm\Mm$. 
\begin{figure}[htbp]
\centering
\includegraphics[width=0.99\columnwidth]{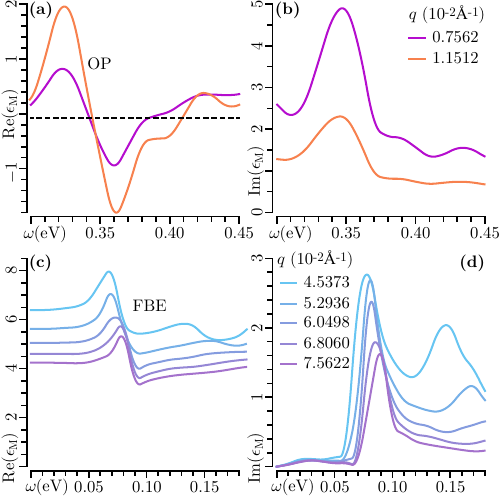}
\caption{\label{fig6} 
(a) Real and (b) imaginary parts of the macroscopic dielectric function $\epsilon_{\mathrm M}(q,\omega)$ across the OP energy range for the two momenta considered in Fig.~\ref{fig5}(b).
(c), (d) Corresponding quantities across the FBE energy range for the momenta considered in Fig.~\ref{fig5}(c).
The dashed horizontal line in (a) indicates $\re[\epsilon_{\mathrm M}(q,\omega)]=0$.
At each momentum, the OP loss maximum lies near the higher-energy zero crossing of $\re[\epsilon_{\mathrm M}(q,\omega)]$.
Across the FBE energy range, by contrast, the real part remains positive while the imaginary part exhibits pronounced
transition-induced structure.
The momentum values and computational settings are identical to those used for the corresponding loss profiles in
Fig.~\ref{fig5}(b) and (c).}
\end{figure}

The dielectric response in the OP energy range at the two momenta considered in Fig.~\ref{fig5}(b) is shown in~Fig.\ref{fig6}(a) and~\ref{fig6}(b). 
At each momentum, the resonant interband response produces two zero crossings of $\re[\epsilon_{\mathrm M}(q,\omega)]$.
The lower-energy crossing lies near the maximum of $\im[\epsilon_{\mathrm M}(q,\omega)]$, where the strong absorptive response suppresses the corresponding enhancement of the loss function.
The OP loss maximum instead occurs near the higher-energy zero crossing, on the descending side of the peak in $\im[\epsilon_{\mathrm M}(q,\omega)]$.
As the momentum transfer increases, this zero crossing shifts toward higher energy together with the OP loss maximum.
The alignment of the loss maximum with a dielectric zero, together with the reduced absorptive response at that energy, supports the identification of this feature as the OP.

The FBE energy range exhibits qualitatively different behavior~[Fig.~\ref{fig6}(c) and~\ref{fig6}(d)].
For all momenta considered in Fig.~\ref{fig5}(c), $\re[\epsilon_{\mathrm M}(q,\omega)]$ remains positive throughout this range.
Although its real part undergoes a pronounced resonant variation, its minima remain well above zero and occur in the presence of an appreciable imaginary part.
The FBE loss maxima instead follow the pronounced structure in $\im[\epsilon_{\mathrm M}(q,\omega)]$, consistent with intra- and interband transitions involving the quasi-flat manifold.
The dielectric response therefore does not provide an unambiguous signature of a well-defined plasmon.
Although some collective contribution cannot be excluded, the feature is conservatively classified as a transition-driven flat-band excitation.
\end{appendix}
\end{document}